\documentclass{iaus}
\usepackage{graphicx}
\newcommand{\arcsec}{\ensuremath{^{\prime\prime}}}

\title[Inflows and Outflows in Nearby AGN] %% give here short title %%
{Inflows and Outflows in Nearby AGN from Integral Field Spectroscopy}

\author[T. Storchi-Bergmann]   %% give here short author list %%
{Thaisa Storchi-Bergmann$^1$}

\affiliation{$^1$Instituto de F\'{i}sica, Universidade Federal do Rio
Grande do Sul, Campus do Vale, CP 15051, 91501-970 Porto Alegre RS, Brazil
\\ Email: {\tt thaisa@ufrgs.br}}

\pubyear{2010}
\volume{267}  %% insert here IAU Symposium No.
\pagerange{119--126}
\jname{Co-Evolution of Central Black Holes and Galaxies}
\editors{B.M.\ Peterson, R.S.\ Somerville, \& T.\ Storchi-Bergmann, eds.}

\begin{document}

\maketitle

\begin{abstract}
I report recent results on the kinematics of the inner few hundred
parsecs (pc) around nearby active galactic nuclei (AGN) at a sampling
of a few pc to a few tens of pc, using optical and near-infrared
(near-IR) integral field spectroscopy obtained with the Gemini
telescopes. The stellar kinematics of the hosts --- comprised mostly of
spiral galaxies --- are dominated by circular rotation in the plane of
the galaxy. Inflows with velocities of $\sim$\,50\,km\,s$^{-1}$ have
been observed along nuclear spiral arms in (optical) ionized gas
emission for low-luminosity AGN and in (near-IR) molecular gas
emission for higher-luminosity AGN. We have also observed gas rotating
in the galaxy plane, sometimes in compact (few tens of pc) disks which
may be fuelling the AGN. Outflows have been observed mostly in ionized
gas emission from the narrow-line region, whose flux distributions and
kinematics frequently correlate with radio flux distributions. Channel
maps along the emission-line profiles reveal velocities as high as
$\sim$\,600\,km\,s$^{-1}$. Mass outflow rates in ionized gas range
from 10$^{-2}$ to 10$^{-3}\,M_\odot$\,yr$^{-1}$ and are 10--100
times larger than the mass accretion rates to the AGN, supporting an
origin for the bulk of the outflow in gas from the galaxy plane
entrained by a nuclear jet or accretion disk wind.

\end{abstract}

\firstsection % if your document starts with a section,
\section{Introduction}

In the present paradigm for nuclear activity in galaxies, the
accretion of gas onto a nuclear supermassive black hole (SMBH) triggers
mechanical and radiative feedback which influences the host galaxies
and intergalactic media of galaxy clusters to which they belong. The
importance of feedback has been recognized in models for co-evolution
of galaxies and black holes (\cite[Hopkins et al. 2005; Di Matteo et
al. 2005]{hopkins05,dimatteo05}), and in solving the ``cooling-flow
problem'' in galaxy clusters (\cite[Rafferty et
al. 2006]{rafferty06}). On galaxy cluster scales, the ``X-ray
cavities'' (\cite[McNamara et al. 2005; Fabian et al. 2006; Allen et
al. 2006]{mcnamara05, fabian06,allen06}) are a signature of strong
feedback from the SMBH of the massive central galaxy. On galactic
scales, the most clear signature of feedback from active galactic
nuclei (AGN) are outflows observed in radio (\cite[Morganti et
al. 2005]{morganti05}), optical (\cite[Storchi-Bergmann et al. 1992;
Komossa et al. 2008]{sb92,komossa08}), UV (\cite[Crenshaw \& Kraemer
2007]{cre07}), and X-rays (\cite[Chelouche \& Netzer
2005]{chelouche05}).  The origin of these outflows seems to be radio
jets and/or accretion disk winds (\cite[Elvis 2000]{elvis00}), most
probably produced by radiation pressure (\cite[Proga 2007; Kurosawa \&
Proga 2008]{proga07,kur08}).

Outflows are a consequence of mass accretion to the SMBH, which
implies transfer of matter to the AGN. Nevertheless, while outflows
are ubiquitous among AGN, inflows are seldom observed. In the present
contribution, I report on a search for inflows in the inner tens to
hundreds of pc around nearby AGN using integral field
spectroscopy. While looking for inflows, we find also outflows and
I will report on their properties as well.

\section{Observations}

We have used integral field spectroscopy (IFS) at the Gemini
telescopes in order to map the gas kinematics in the inner
$\sim\!300$\,pc around nearby AGN. The final product of the IFS
observations are ``datacubes'', which have two spatial dimensions ---
allowing the extraction of images covering a range of wavelengths ---
and one spectral dimension --- allowing the extraction of spectral
information of each spatial element.

In the optical, we have used the Integral Field Unit of the Gemini
Multi-Object Spectrograph (IFU-GMOS), which has a field-of-view of
$3.\!''5\ \times 5$\arcsec\ in one-slit mode or
5\arcsec\,$\times$\,7\arcsec\ in two-slit mode at a sampling of
$0.\!''2$ and angular resolution (dictated by the seeing) of
$0.\!''6$, on average. The resolving power is $R \approx 3000$.
% and we observed in the wavelength range 7500--9300\AA.

In the near-infrared (near-IR) we have used the Near-Infrared Integral
Field Spectrograph (NIFS) together with the adaptative optics module
ALTAIR (ALTtitude conjugate Adaptive optics for the InfraRed), which
delivers an angular resolution of $\sim\!0.\!''1$. The
field-of-view is 3\arcsec\,$\times$\,3\arcsec\ at a sampling of
$0.\!''04 \times 0.\!''1$ and the spectral resolution is
$R\approx 5000$. We have also used the IFU of the Gemini
Near-Infrared Spectrograph (GNIRS) with a field-of-view of
3\arcsec\,$\times$\,5\arcsec, and resolving power of $R \approx 5900$
covering the $J$, $H$, and $K$ bands.

\section{Inflows}

Nuclear spirals --- on scales of hundred parsecs --- are frequently
observed around AGN in images obtained with the {\em Hubble Space Telescope
(HST)} (\cite[Martini et al. 2003]{martini03}). \cite[Martini \& Pogge
(1999)]{martini99} have shown that these spirals are not
self-gravitating and may be the channels through which matter is being
transferred to the nucleus to feed the AGN. This interpretation is
supported by models (\cite[Maciejewski 2004]{maciejewski04}) and by
results such as those from \cite[Prieto et al. (2005)]{prieto05} for
the galaxy NGC\,1097 and from \cite[Sim\~oes Lopes et
al. (2007)]{sl07} for a larger sample of galaxies. The latter authors
have built ``structure maps'' using images obtained with the {\em HST}
Wide-Field and Planetary Camera 2 (WFPC2) through the F606W filter of
a sample of AGN and a matched sample of non-active galaxies. The
structure maps revealed dusty nuclear spirals in all early-type AGN
hosts, but in only $\sim$\,25\% of the non-AGN, indicating that these
spirals are strongly linked to the nuclear activity and map the matter
in its way to feed the SMBH at the nucleus.
%More results on the sample of  \cite[Sim\~oes Lopes et al. (2007)]{sl07} on the basis of Spitzer observations are discussed in the work  by this author in the present volume.
Although the correlation between the nuclear spirals and the nuclear activity is strong, it is based only on morphology. In order to check if there are indeed inflows along the nuclear spirals, we began a project to map such inflows using the Gemini integral field spectrographs.

\subsection{Optical Observations}

%\begin{figure}[b]
% \vspace*{-2.0 cm}
%\begin{center}
% \includegraphics[width=5in]{quadrof346.eps} 
% \vspace*{-1.0 cm}
%\caption{Bottom: third, fourth and sixth eigen-spectra. Top: corresponding tomograms. The anti-correlation of the emission lines, visible in the eigen-spectra indicate anti-correlated blueshifts and redshifts, which suggest rotation around the nucleus (see also next Figure).}
%\label{f346}
%\end{center}
%\end{figure}

\begin{figure}[b]
% \vspace*{-2.0 cm}
\begin{center}
\includegraphics[width=4in]{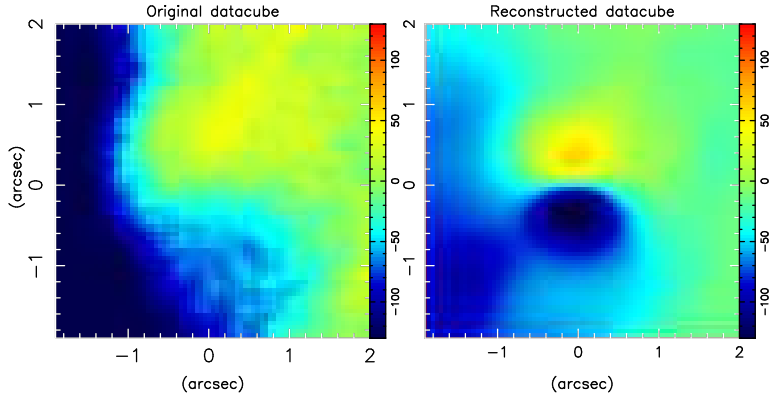} 
% \vspace*{-1.0 cm}
\caption{Left: Gas kinematics obtained from the centroid velocities of
the [N{\,\sc ii}]\,$\lambda$6584 emission line in the datacube of
M\,81. Right: Gas kinematics obtained from the same velocticies
measured in the reconstructed datacube after the PCA analysis,
considering only the contribution of 3 eigen-spectra which reveal a
compact ($\sim\!8$\,pc radius) rotating disk around the
nucleus. Velocities are color-coded as in the bar to the right, in
units of km\,s$^{-1}$.}
\label{quadrovel}
\end{center}
\end{figure}

\noindent\paragraph{\underline {\it Inflows in NGC\,1097 and NGC\,6951.}} The GMOS-IFU was
used to map the gas kinematics in the inner
7\arcsec\,$\times$\,15\arcsec\ of the LINER/Seyfert 1 galaxy NGC\,1097
(\cite[Fathi et al. 2005]{fathi06}), corresponding to the inner
500\,pc\,$\times$\,1\,kpc of the galaxy. We have fitted Gaussians to
the H$\alpha$ and [N\,{\sc ii}]\,$\lambda$6584 emission lines  and find
that the kinematics is dominated by rotation in the galaxy plane, but
with non-circular motions superimposed. In order to isolate these
motions, we fitted a circular rotating disk model to the velocity
field and subtracted the model from the measured velocities. The
residual velocity field was spatially correlated with a nuclear spiral
structure, supporting previous suggestions (\cite[Prieto et
al. 2005]{prieto05}) that these spirals are indeed associated with
inflows towards the nucleus. Further analysis of the gas kinematics in
the nuclear region of NGC\,1097 was recently published by \cite[Davies
et al. (2009)]{davies09}, confirming the inflows along the nuclear
spirals. Similar results were obtained for another LINER galaxy,
NGC\,6951 (\cite[Storchi-Bergmann et al. 2007]{sb07}).  The streaming
motions along the nuclear spirals have velocities of
$\sim\!50$\,km\,s$^{-1}$ and the estimated mass inflow rate in
ionized gas is $\sim$\,10$^{-3}\,M_\odot$\,yr$^{-1}$. Coincidently,
this mass inflow rate is of the order of the mass accretion rate to
the active nucleus --- under the assumption that the AGN luminosity is
extracted from the mass accretion rate in the prescription of a
radiatively inneficient accretion flow, which seems to apply to such
nuclei (\cite[Nemmen et al. 2006]{nemmen06}). The actual mass inflow
in neutral and molecular gas is probably much larger than that in
ionized gas estimated here.  \\

\noindent\paragraph{\underline{\it Inflows in M\,81.}} We have also mapped the gas
kinematics in the inner $7\arcsec\times15\arcsec$ of the nearby
LINER/Seyfert 1 galaxy M\,81. Due to its proximity, the region covered
at the galaxy is only $\sim\!120\,{\rm pc} \times 250$\,pc, at a spatial
resolution of $\sim\!10$\,pc. At these small galactic scales, the
gas velocity field does not show a clear rotation pattern, and we used
a particular method to isolate non-circular motions: we obtained the
stellar velocity field using the ``penalized pixel fitting technique
(pPXF)'' of \cite[Capellari \& Emsellem (2004)]{capellari04}, and
subtracted the stellar velocity field from that of the gas. The
results were blueshifts in the gas kinematics on the far side of the
galaxy and redshifts on the near side, suggesting inflows along the
galaxy minor axis. A further technique we are exploring is the use of
principal component analysis (PCA) applied to the datacube. This
method is described by \cite[Steiner et al. (2009)]{steiner09} and
by Steiner et al.\ in these proceedings. The
result is the separation of the information in ``eigen-spectra,'' which
reveal spatial correlations and anti-correlations in the emission-line
data.
%Then asociated ÒtomogramsÓ map the spatial distribution of the eigenvectors. 
%A preliminary result of its application to the inner $5\arcsec\,\times\,5\arcsec$ (85\,$\times$\,85\,pc$^2$) of the M81 datacube is illustrated in Figs.\,\ref{f346} and  \ref{quadrovel}. Fig.\,\ref{f346} shows that the third, fourth and sixth eigen-spectra show anti-correlated peaks in the H$\alpha$ and [N{\,\sc ii}]$\lambda$6584\AA\ emission lines, which can be interpreted as resolved rotation. 
A preliminary result of the application of this technique to the inner
$5\arcsec\times5\arcsec$ (85\,$\times$\,85\,pc$^2$) of the M\,81
datacube is illustrated in Figure \ref{quadrovel}, which shows a
comparison between the kinematics derived from the datacube and that
from a ``reconstructed'' datacube (\cite[Steiner et
al. 2009]{steiner09}) using only 3 eigen-spectra. The latter reveals
what seems to be a compact rotating disk ($\sim$\,8\,pc radius) around
the nucleus. More results on M\,81 are discussed 
by Schnorr M\"uller et al.\ in these proceedings.

\subsection{Near-Infrared Observations}

\noindent \paragraph{\underline {\it Inflows in NGC\,4051.}} We have looked for inflows
toward AGNs in the near-IR using NIFS with the adaptative optics module
ALTAIR at the Gemini North Telescope. We have mapped the stellar and
gas kinematics in the inner $\sim$\,100\,pc of the Seyfert galaxy
NGC\,4051 from $K$-band spectra. Channel maps along the 
H$_2\,\lambda\,2.12\,\mu$m emission-line
profiles show blueshifts on the far side and
redshifts on the near side of the galaxy along nuclear spirals which
can be interpreted as inflows towards the nucleus if the gas is in the
plane of the galaxy (\cite[Riffel et al. 2008]{riffel08}).  \\
\begin{figure}[t]
\centering
\begin{minipage}{0.55\linewidth}
\hspace*{-1.0 cm}
\includegraphics[height=.5\textheight]{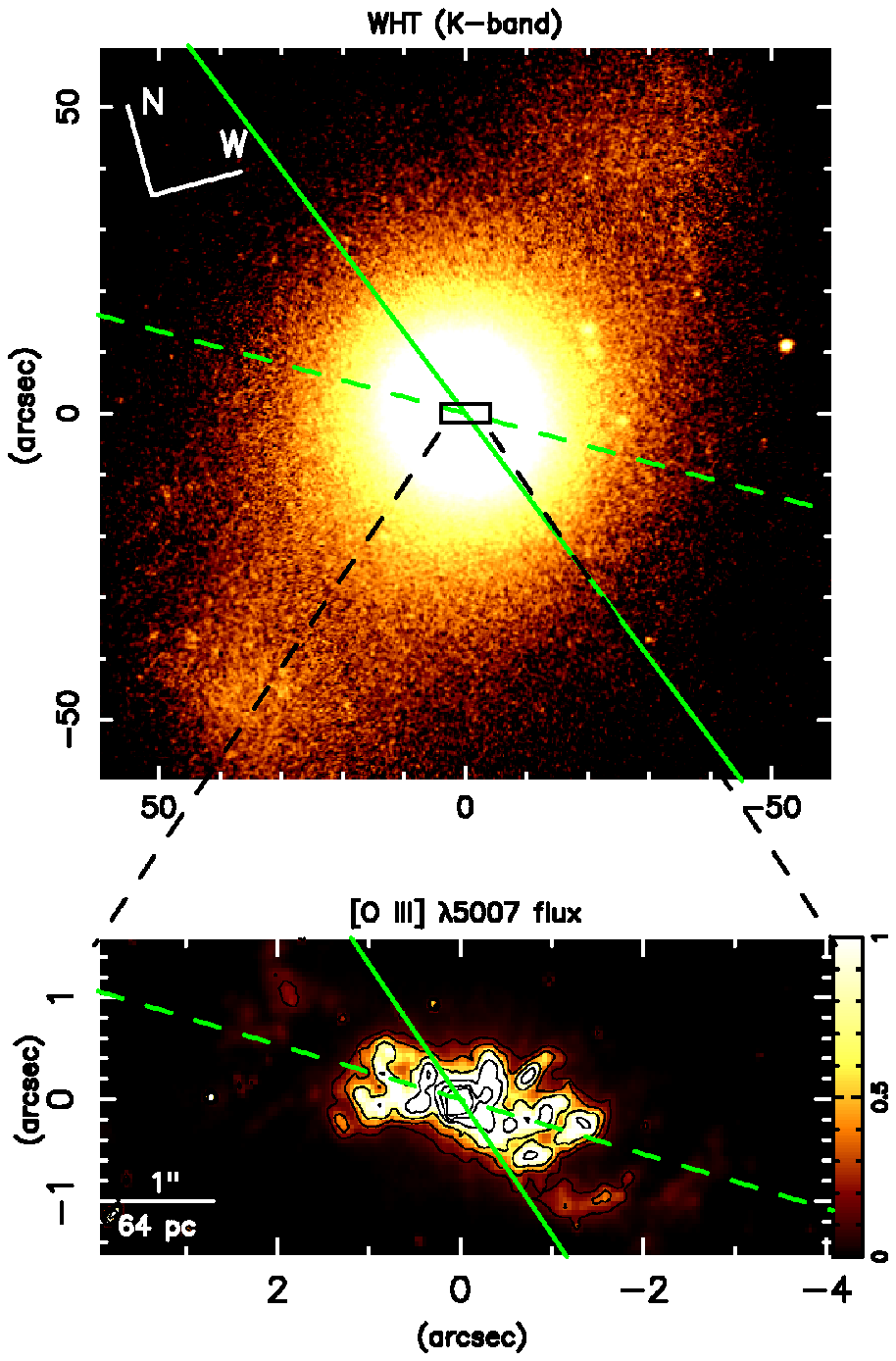}
\end{minipage}
\begin{minipage}{0.35\linewidth}
\hspace*{-1.5 cm}
\includegraphics[height=.5\textheight]{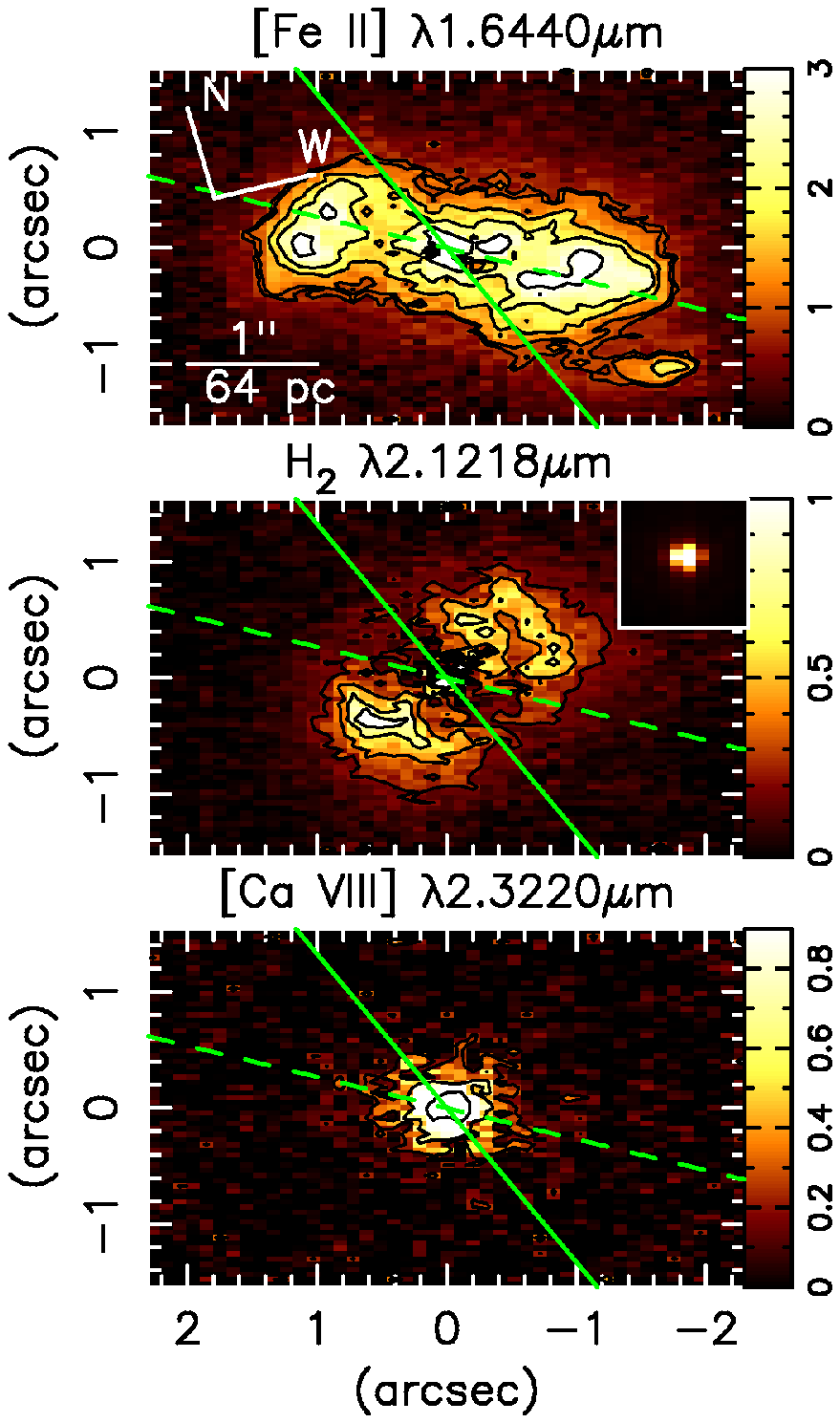}
\end{minipage}
\caption{Left top: $K$-band image of the central region of NGC\,4151
obtained with the William Herschel Telescope. The continuous line
shows the orientation of the major axis of the galaxy, and the dashed
line that of the NLR bi-cone. The rectangle shows the region covered
by the NIFS observations. Left bottom: {\em HST} [O\,{\sc iii}]\,$\lambda$5007
narrow-band image of the NLR in the NIFS field-of-view. 
Right: Intensity maps in the inner 3$\times$5\,arcsec$^2$. Top: [Fe\,{\sc
 ii}] intensity map; middle: H$_2$ intensity map; bottom: a coronal
 line intensity map (\cite[Storchi-Bergmann et al. 2009]{sb09}). }
\label{n4151_flux}  
\end{figure}

\begin{figure}[b]
%\vspace*{-1.0 cm}
\begin{center}
\hspace*{-1.0 cm}
\includegraphics[width=5.3in]{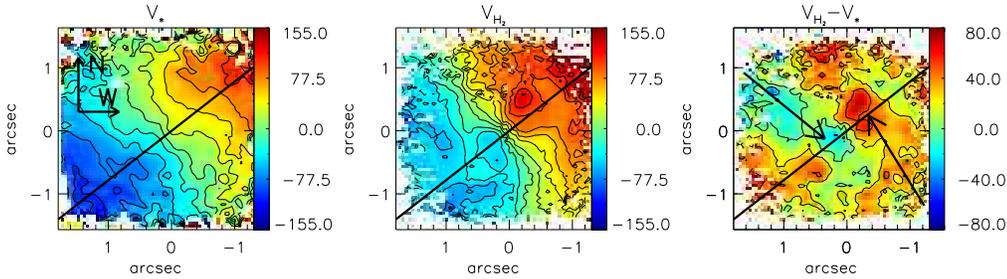} 
\caption{Kinematics of the inner 700\,$\times$\,700\,pc$^2$ of
Mrk\,1066. From left to right: stellar velocity field, molecular gas
velocity field (from centroid of the H$_2\,\lambda\,2.1218\mu$m
emission-line) and residual between the molecular and stellar velocity
fields. Note the compact rotating disk in H$_2$ emission. The black
line shows the line of nodes derived from the stellar kinematics,
while the arrows show the spiral arms, which become noticeable only in
the residual map. Velocities are color-coded as indicated in the bar
to the right in units of km\,s$^{-1}$.}
 \label{mrk1066_inflow}
\end{center}
\end{figure}

\noindent \paragraph{\underline {\it Inflows in NGC\,4151.}} We have also mapped the gas
excitation and kinematics with NIFS in the inner $\sim$\,
200$\times$500\,pc$^2$ of the Seyfert galaxy NGC\,4151 at a spatial
resolution of $\sim$\,8\,pc (\cite[Storchi-Bergmann et al. 2009,
2010]{sb09, sb10}). We have found distinct flux distributions for the
ionized, molecular, and coronal gas, as illustrated in
Figure \ref{n4151_flux}: while the ionized flux distributions follow the
ionization bicone previously observed in the [O\,{\sc
iii}]\,$\lambda$5007 emission line --- which characterizes the
narrow-line region (NLR) of this galaxy --- the molecular gas avoids
the bicone region, and the coronal gas emission is barely resolved.

Besides the flux distribution, the kinematics of the molecular and
ionized gas is also distinct: the molecular gas shows only rotation,
and is thus probably confined to the galactic plane, while the ionized
gas shows both rotation and outflows. The ionized gas emission is not
restricted to the bicone, where the dominant kinematics is outflow
from the nucleus, but shows also emission from the galactic plane
within $\sim\,$65\,pc from the nucleus, where the gas kinematics is
dominated by rotation (similarly to the molecular gas). Thus, although
we could not map actual inflows in NGC\,4151, the circumnuclear gas
rotating in the plane is probably the mass reservoir to feed the
SMBH. In support to this idea, a previous study of the H\,{\sc i} gas
kinematics has found inflows along the large scale bar (\cite[Mundell
et al. 1999]{mundell99}), oriented approximately along the galaxy
minor axis. This flow could have built the molecular gas reservoir we
have obseved within a few tens of parsecs from the nucleus along the
galaxy minor axis, which has a similar orientation to that of the
large scale bar.

The NIFS observations of NGC\,4151 have also revealed the presence of
an unresolved red nuclear continuum, well reproduced by a black body
with temperature $T=1340$\,K (\cite[Riffel et al. 2009]{riffel09}),
and identified with the hottest part of the dusty torus postulated by
the unified model of AGNs (\cite[Antonucci \& Miller
1985]{antmi85}). This torus may originate in inflows from the
molecular gas reservoir described above.  \\

\noindent\paragraph{\underline{\it Inflows in Mrk\,1066.}} We have used NIFS to map also
the stellar and gas kinematics in the inner $\sim\!\,350$\,pc of the
Seyfert 2 galaxy Mrk\,1066, using $J$ and $K$-band
spectroscopy. Preliminary results for the stellar and molecular
(H$_2$) gas velocity fields are shown in
Figure \ref{mrk1066_inflow}. The blueshifts on the far side of the
galaxy and redshifts on the near side observed in the H$_2$ residual
map suggest inflows towards the center along nuclear spiral
arms. These arms seem to feed a rotating disk with radius of
$\sim\!\,100$\,pc around the nucleus. Further results on the
kinematics of the nuclear region of Mrk\,1066 are discussed
by Riffel et al. (these proceedings).

\section{Outflows}

\subsection{Optical Observations}

\noindent\paragraph{\underline{\it Outflows in NGC\,2273, NGC\,3227, NGC\,4051 and NGC\,3516.}} 
We have obtained data on the stellar and gaseous kinematics of the inner
$\sim\!400$ pc of 6 Seyfert galaxies using the GMOS--IFU in the
wavelength range $\sim\!7500-9300$\AA\ (\cite[Barbosa et
al. 2009]{barbosa09}). The stellar kinematics was obtained using the
calcium triplet at $\approx\,$8500\AA\ and the gas kinematics using
the [S\,{\sc iii}]$\lambda$9069\AA\ emission line. The stellar
kinematics is dominated by circular rotation in the galaxy plane. The
gaseous kinematics shows also a rotational component, intepreted as
due to gas rotating in the plane as the stars. In these four galaxies,
the gas kinematics shows, in addition, blueshifts and
redshifts due to outflows from the nucleus, whose velocities (derived
from the centroids of the emission lines) reach
$\sim\!\,200$\,km\,s$^{-1}$ at a few hundred parsecs from the
nucleus. The outflows are spatially associated with the radio
structure and both the radio flux maps and the [S\,{\sc\,iii}] flux
and centroid velocity maps show discrete components (knots of
emission), which we interprete as due to intermitent ejection of
plasma which compresses the surrounding interstellar medium, driving
the observed outflows. This interpretation is supported by the
observation of an increase in the gas velocity dispersion in the
regions surrounding the radio knots.

Channel maps along the [S\,{\sc iii}] emission line show that the the
highest velocities are observed close to the nucleus, contrary to what
is observed in the centroid velocity maps, which show the highest
velocities away from the nucleus. We attribute this difference to the
fact that the centroid velocity probes the brightest emission: in the
vicinity of the nucleus the brightest component is the one originating
in the galactic disk (with velocities close to systemic), while away
from the nucleus the brightest component is the outflowing one. As a
consequence, the centroid velocities show an apparent increase in
velocity that mimics acceleration along the NLR. In the channel maps,
we see high velocity gas in the nucleus, showing that the outflow does
not leave the nucleus at zero velocity. We interpret this high
velocity gas as an outflowing wind from the AGN or ambient gas more
directly interacting with this wind. As this highest velocity gas
moves away from the nucleus, it pushes and accelerates the gas from
the disc.

We have estimated mass outflow rates in the range
10$^{-3}$--10$^{2}\,M_\odot$\,yr$^{-1}$. These are approximately 10
times the accretion rates of their respective AGN, indicating that the bulk of the
outflowing gas is entrained from the galaxy ISM (\cite[Veilleux et
al. 2005]{veilleux05}). We have also estimated that the power of the
outflow is about 10$^{-4}$ times the bolometric luminosity.

\subsection{Near-Infrared Observations}

 \begin{figure}[b]
 \begin{center}
\includegraphics[scale=0.6]{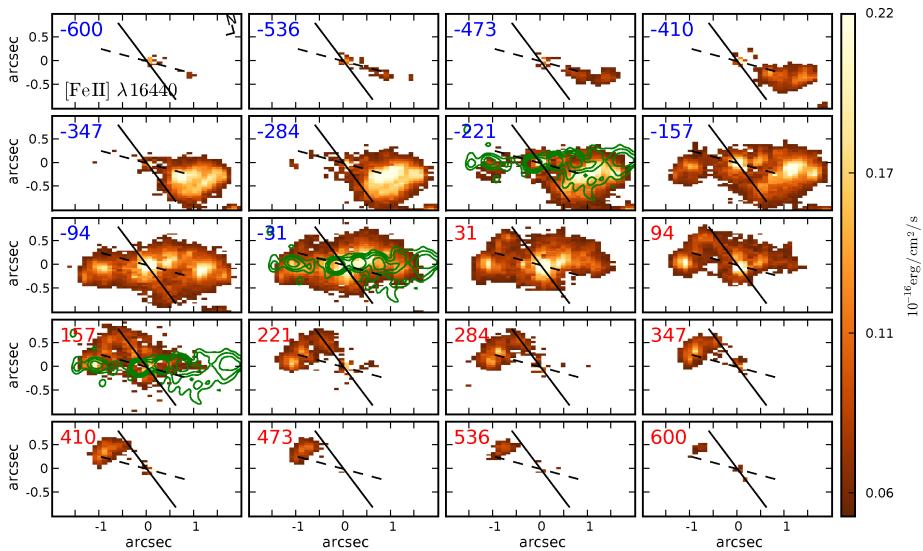}
\caption{Channel maps obtained by integrating the flux within velocity
bins of $63$\,km\,s$^{-1}$ along the [Fe{\, \sc
ii}]\,$\lambda 1.64\mu$m emission-line profile from the NLR of
NGC\,4151. The numbers in the upper left corner of each panel are the
central velocity of the bin, in km\,s$^{-1}$ relative to systemic. The
continuous line shows the orientation of the galaxy major axis and the
dashed line shows the orientation of the NLR bicone. Contours are
from a radio MERLIN image. }\label{channel_n4151}
\end{center}
\end{figure}

\noindent\paragraph{\underline {\it Outflows in NGC\,4151.}} As discussed above, NIFS
observations of NGC\,4151 show that part of the ionized gas
surrounding the nucleus is orbiting in the plane and may comprise the
fuelling flow to the AGN. Part of the gas is also outflowing, with
centroid velocities reaching up to 400\,km\,s$^{-1}$ away from the
nucleus. Channel maps along the emission-line profiles reveal even
higher velocities (up to $\sim\!600$\,km\,s$^{-1}$) which are
observed close to the nucleus, as illustrated in
Figure \ref{channel_n4151} (\cite[Storchi-Bergmann et al. 2010]{sb10}),
and do not support acceleration along the NLR as suggested by previous
authors.

The axis of the bicone observed in the optical and near-IR flux maps
shows a distinct orientation from that of a nuclear radio jet. As the
radio jet probably originates in the ``funnel'' of the accretion disk,
the origin of the emitting gas cannot be the same. The biconical
outflow probably originates from an accretion disk wind, and the
distinct orientation suggests that the accretion disk is warped. The
channel maps close to zero velocity (systemic velocity) show
nevertheless some correlation between the emitting gas intensities and
the radio knots, as can be observed in Figure \ref{channel_n4151}. We
interpret this correlation as due to interaction of the radio jet with
gas from the galaxy plane, which is almost in the plane of the sky and
thus have observed velocities close to zero. We estimated a mass
outflow rate of $\sim\! 1 M_\odot$\,yr$^{-1}$ along each cone,
which exceeds the inferred black hole accretion rate to the AGN by a
factor of $\sim\,100$. The kinetic power of the outflow as measured
from the emission-line intensities and velocities is
$\sim\!3\,\times\,10^{-3}$ times the bolometric luminosity.  \\

\noindent\paragraph{\underline{\it Outflows in ESO\,428--G14,
NGC\,7582, Mrk\,1066, Arp\,102B, and SDSS\,J0210--0903.}} We \\
have found a
strong correlation of the ionized gas flux distributions and
kinematics with the radio-flux distributions in the Seyfert galaxies
ESO\,428--G14 (\cite[Riffel et al. 2006]{riffel06}) and
Mrk\,1066. Further results for the latter galaxy are discussed in the
contribution by 
Riffel \& Storchi-Bergmann (these proceedings). Outflows have been found
also in the nuclear region of NGC\,7582 (\cite[Riffel et
al. 2009a]{riffel09a}). The radio galaxy Arp\,102B shows an outflow
correlated with the radio flux distribution
(Couto, these proceedings), Finally, the gas kinematics of the
post-starburst quasar SDSS\,J0210--0903 is discussed 
by Sanmartin, Storchi-Bergmann, \& Brotherton (these proceedings).

\section{Summary and Conclusions}

I have discussed a number of studies of the gas kinematics in the
inner few hundred parsecs around nearby AGN using integral field
spectroscopy. These data have been used to search for gas inflows
which feed the nuclear SMBH. In doing so, we also mapped outflows in
ionized emitting gas which are more easily observable than inflows.\\

\noindent\paragraph{\underline {\it Inflows.}} In the optical, we were able to map inflows
observed in ionized gas around the low-luminosity AGNs NGC\,1097,
NGC\,6951, and M\,81. In the first two galaxies, we found streaming
motions towards the nucleus on scales of hundred parsecs along spiral
arms. In the latter, we used PCA analysis to unveil a compact rotating
disk with radius $\sim\!8$\,pc, which may be fuelling the AGN. In
more luminous AGN, we were successful in finding similar kinematics in
molecular gas emission in the near-IR. The H$_2$ gas kinematics reveal
both streaming motions towards the nucleus and compact rotating
disks. The mass inflow rates in ionized gas are of the order of the
nuclear accretion rate (derived from the luminosity of the AGN), while
those in molecular gas are much smaller. In both cases, we conclude
that we are only seeing the ``hot skin'' (ionized/excited by the AGN)
of a much more massive inflow of colder molecular and neutral gas. In
the optical, we have also observed emission from ionized gas in
rotation around the nucleus, and conclude that this gas may be part of
the mass reservoir being used to feed the AGN.\\

\noindent\paragraph{\underline{\it Outflows.}} 
%We have found outflows around most AGN. 
%Even in the LINERs NGC\,6951 and M\,81 there are signatures of mild outflows in ionized gas (\cite[Storchi-Bergmann et al. 2007, Schnorr M\"uller et al. in preparation]{sb07}). 
Outflows are observed mostly in ionized gas emission, and the centroid
velocity maps usually show larger velocities --- reaching typically
200--400\,km\,s$^{-1}$ --- away from the nucleus (hundred of parsecs
scales), suggesting acceleration along the NLR. Nevertheless, channel
maps extracted along the emission-line profiles reveal high velocities
(up to $\sim\!600$\,km \,s$^{-1}$) close to the nucleus and do not
support acceleration. We suggest that the apparent acceleration
observed in the centroid velocity maps is due to the fact that the
brightest component --- which is probed by these maps --- is the disk
emission near nucleus (which has low velocity) but is the outflow
emission outwards (which has higher velocity). The mass outflow rates
along the NLR are in the range 10$^{-3}-10^{-2}\,M_\odot$\,yr$^{-1}$
which are 10 to 100 times the mass accretion rate to the SMBH. This
result supports previous claims (e.g., \cite[Veilleux et
al. 2005]{veilleux05}) that the origin of the NLR gas is entrainment
of the galaxy ISM by a radio jet or accretion disk wind. We have
estimated the kinetic power of the outflows as $\sim\,10^{-4}-10^{-5}$
times the bolometric luminosity.

%\noindent{\it Comparison betwee Seyfert and LINERS}
%Nemmen et al. in these Proceedings have derived accretion rates and jet power from fits of the SED of low-luminosity AGN: L/LEdd= 10-3 Ð 10-2 for Sy, 10-8 Ð 10-4 for LINERs ? lower luminosity AGN;(dM/dt)acc= 10-4-10-2 for both Sy and LINERs; Pjet/Lbol  = 10-3 for Sy, 5 for LINERs ? powerful jets in LINERs

\end{document}